\documentclass[12pt,letterpaper,a4paper]{article}
\usepackage[includeheadfoot,
            marginratio={1:1,2:3},
            width=455pt,
            height=786pt,]{geometry}
\usepackage{amsmath}
\usepackage{amsfonts}
\usepackage{amssymb}
\usepackage{graphicx}
\usepackage{cancel}

\usepackage{empheq}
\usepackage{color}
\usepackage{hyperref}

\usepackage{float}
\restylefloat{table}
\restylefloat{figure}

\newcommand{\nc}{\newcommand}
\nc{\beq}{\begin{equation}}
\nc{\eeq}{\end{equation}}
\nc{\bea}{\begin{eqnarray}}
\nc{\eea}{\end{eqnarray}}

\def\ov{\overline}

\allowdisplaybreaks[2]
\numberwithin{equation}{section}

\begin{document}

\vspace{1.5cm}
\begin{center}
{\LARGE
Some inequalities bridging stringy parameters and \vskip0.2cm cosmological observables}
\vspace{0.4cm}
\end{center}

\vspace{0.35cm}
\begin{center}
Anupam Mazumdar$^\dag$ and Pramod Shukla$^\ddag$
\end{center}

\vspace{0.1cm}
\begin{center}
\emph{
$^\dag$Consortium for Fundamental Physics, Physics Department, Lancaster University, LA1
4YB, United Kingdom\\
\vspace{0.4cm}
$^\ddag$Universit\'a di Torino, Dipartimento di Fisica and I.N.F.N. - sezione di Torino \\
Via P. Giuria 1, I-10125 Torino, Italy \footnote{From October 1, 2015, the address of Pramod Shukla has been changed to {\it ICTP, Strada Costiera 11, Trieste 34151, Italy}, with the email: shukla.pramod@ictp.it.}
}

\vspace{0.2cm}

\vspace{0.5cm}
\end{center}

\vspace{1cm}

\vspace{1cm}
%%%%%%%%%%%%%%%%%%%%%%%%%%%%%%%%%%%%%%%%%%%%%%%
%%%%%%%%%%%%%%%%%%%%%%%%%%%%%%%%%%%%%%%%%%%%%%%
%%%%%%%%%%%%%%%%%%%%%%%%%%%%%%%%%%%%%%%%%%%%%%%
%%%%%%%%%%%%%%%%%%%%%%%%%%%%%%%%%%%%%%%%%%%%%%%
%%%%%%%%%%%%%%%%%%%%%%%%%%%%%%%%%%%%%%%%%%%%%%%
%%%%%%%%%%%%%%%%%%%%%%%%%%%%%%%%%%%%%%%%%%%%%%%
%%%%%%%%%%%%%%%%%%%%%%%%%%%%%%%%%%%%%%%%%%%%%%%
%%%%%%%%%%%%%%%%%%%%%%%%%%%%%%%%%%%%%%%%%%%%%%%

\begin{abstract}
By demanding the validity of an effective field theory description during inflation, in this note we derive some peculiar inequalities among the three interesting stringy and cosmological parameters, namely the tensor-to-scalar ratio ($r$), the string coupling ($g_s$) and the compactification volume (${\cal V}$). In deriving these inequalities, we explicitly demand that  the inflationary scale and the Hubble parameter during inflation are well below the Kaluza-Klein (KK) mass scale, string scale, and the four dimensional Planck mass. For the inflationary models developed within the framework of type IIB orientifold comapctification, we investigate the regions of parameters space spanned by the three parameters $(r, g_s, {\cal V})$ by satisfying our inequalities, and we find that the same can reduce the size of available parameter space quite significantly. Moreover, we comment on obtaining further constraints on the parameters by comparing gravitino mass ($m_{3/2}$) with the Hubble scale ($H$), which also provides a lower bound on tensor-to-scalar ratio ($r$), for the cases when $m_{3/2} <H$. We also illustrate the outcome of our bounds in some specific class of string(-inspired) models. 
\end{abstract}

\clearpage

\tableofcontents

%%%%%%%%%%%%%%%%%%%%%%%%%%%%%%%%%%%%%%%%%%%%%%%
%%%%%%%%%%%%%%%%%%%%%%%%%%%%%%%%%%%%%%%%%%%%%%%
%%%%%%%%%%%%%%%%%%%%%%%%%%%%%%%%%%%%%%%%%%%%%%%
%%%%%%%%%%%%%%%%%%%%%%%%%%%%%%%%%%%%%%%%%%%%%%%
%%%%%%%%%%%%%%%%%%%%%%%%%%%%%%%%%%%%%%%%%%%%%%%
%%%%%%%%%%%%%%%%%%%%%%%%%%%%%%%%%%%%%%%%%%%%%%%
%%%%%%%%%%%%%%%%%%%%%%%%%%%%%%%%%%%%%%%%%%%%%%%
%%%%%%%%%%%%%%%%%%%%%%%%%%%%%%%%%%%%%%%%%%%%%%%

\section{Introduction and Motivation} 
The primordial inflation is one of the best known paradigms for explaining the large scale structures in the universe, and the 
origin of temperature anisotropy in the cosmic microwave background (CMB) radiation~\cite{Planck:2013jfk}. However, it is a challenge to build 
a model of inflation which can simultaneously explain the observables for CMB, matter perturbations, and predict the right thermal
history of the universe from the end of inflation until now, for a review see~\cite{Mazumdar:2010sa}.

Moreover, it is equally challenging to embed inflation correctly within an effective field theory (EFT) due to the presence of multiple-scales 
in the ultraviolet (UV) physics~\cite{Chialva:2014rla}. This problem has become prominent due to the claim of a discovery of primordial
gravitational waves and the large tensor-to-scalar ratio by BICEP2~\cite{Ade:2014xna}. Although the previous claims of BICEP2 have been diluted in the follow-up joint investigations of BICEP2/Keck Array and PLANCK results \cite{Ade:2015tva,Ade:2015lrj}, one can still hope that in future large tensor-to-scalar ratio could still be detectable, such as  $r\simeq {\cal O}(0.05)$, or so. Moreover, the most recent observational  constraint  by  BICEP2  and  the  Keck  Array  on  the  tensor-to-scalar  ratio has been argued to be at $r <0.07$ at 95 \% confidence \cite{Array:2015xqh}.  

Although there exists models of inflation where  such values of 
$r\sim {\cal O}(0.05)$ can be obtained well within sub-Planckian excursion of the inflaton field, see~\cite{Hotchkiss:2011gz,Chatterjee:2014hna}.
Indeed such models can be embedded within supergravity models of inflation~\cite{Choudhury:2014sxa}, and it is compatible with the effective field 
theory treatment, since the inflaton VEV is always bounded by the  $4$-dimensional (4-D) Planck mass, $M_p = (8 \pi G_N)^{-1/2} = 2.4 \times 10^{18}$~GeV.

The same cannot be said apriori if there are more than one scales in the problem. 
Inflationary models developed in  four dimensional effective field theory framework, which are obtained after compactifying the ten dimensional superstring theories on suitable class of (Calabi Yau) manifolds, contain many scales besides the (4-D) Planck mass,  for a review on inflation within string theory, see~\cite{Burgess:2013sla, Westphal:2014ana, Baumann:2014nda}. These scales are: string scale, $m_s$, the lightest Kaluza-Klein (LKK) mass scale, $m_{KK}$, and the winding mode, $m_W$. Typically, there is a hierarchy in these scales, which is given by: $$m_{KK} < m_{s} < m_W< M_{p}.$$ In order to have a successful period of
 inflation, i.e. $50-60$ e-foldings of inflation, one would also have to demand that the Hubble expansion rate during inflation, $H_{\rm inf}$, follows:
\bea\label{Bound}
H_{\rm inf} \ll m_{KK}< m_{s} < m_W< M_{p}\,.
\eea
A simple reason for such a stringent demand arises from the validity of an EFT at the lowest order. There are obvious consequences if we had to violate this 
bound. If $H_{\rm inf} \geq m_{KK}$, we would end up exciting not only the LKK, but also the tower of KK modes during inflation.  This will immediately backreact
into the original potential and might alter the predictions. Although, if somehow inflation had triggered then these heavy states would be washed 
away during inflation, but again they would be excited abundantly after the end of inflation, via non-perturbative mechanisms~\cite{Kofman:2005yz,Frey:2005jk}. 
In some cases, the LKK could be absolutely stable and would overclose the universe prematurely by such non-perturbative excitations~\cite{Kofman:2005yz,Frey:2005jk,Cicoli:2010ha,Cicoli:2010yj, Chialva:2012rq}.

This would be a mere catastrophe for embedding inflation within string theory. Describing inflation within 4-D when KK-modes, winding modes are all excited is beyond the scope of current understanding, because there 
would be inherent stringy corrections to the inflaton potential arising from higher order string couplings, $g_s$-corrections \cite{Berg:2005ja,Berg:2007wt,Cicoli:2007xp} and $\alpha'$-corrections \cite{Becker:2002nn}, which would induce 
higher derivative corrections to the gravitational sector~\footnote{For a recent study beyond the two-derivative approximation, see \cite {Ciupke:2015msa,Ciupke:2016agp} in which the higher order $F^4$-corrections induced from $\alpha^\prime$ have been studied in type IIB orientifold framework.}, 
which cannot be computed so easily in a  time dependent background. In many cases, if the scale of inflation is higher than the compactification scale, it would be very hard to 
understand the complicated dynamics  in a de-compactification limit %- how and why the $3$ spatial dimensions expand, while $6$ dimensions shrink~
\cite{Brandenberger:1988aj,Danos:2004jz}. To avoid all these constraints we would require the inflationary potential, $V_{\rm inf}$:
\bea\label{Bound-1}
V_{\rm inf}^{1/4} \sim ( 3H_{\rm inf}^2M_p^2)^{1/4} \ll m_{KK}\,.
\eea
The aim of this paper is very simple.  Given all these constraints, if we wish to be within a valid EFT regime, i.e. by following Eqs. (\ref{Bound})-(\ref{Bound-1}),
could we then obtain a simple bound on the value of tensor-to-scalar ratio ($r$), with the help of string coupling ($g_s$), and the compactifticaion volume
(${\cal V}$) in a rather model independent way. In order to illustrate our point,  we will discuss all the relevant scales and their hierarchies  in the framework of type IIB orientifold compactification in which moduli stabilization has been studied with a fairly better understanding in the two well known schemes, namely the KKLT \cite{Kachru:2003aw} and the LARGE volume scenarios \cite{Balasubramanian:2005zx}. 
 
\section{Analytic expressions of various scales}
In the section, we will collect the relevant expressions of various scales involved via considering a setup of type IIB superstring theory compactificed on a Calabi Yau orientifold.

\subsection{String scale $m_s$} 
By following the conventions \cite{Baumann:2014nda, McAllister:2008hb, Dienes:2002ze} as $ \hbar = c =1$, and the string length $l_s = \sqrt{\alpha^\prime}$, which subsequently sets the string mass as $m_s = l_s^{-1}$, one can write the effective 4-D type IIB supergravity action in the string frame, within
no warping limit, see~\cite{Baumann:2014nda}:
\bea
\label{eq:10dto4d}
S_{IIB} \approx \frac{1}{(2 \pi)^7 \, (\alpha^\prime)^4 \, g_s^2} \int d^4 x \, \sqrt{- {g}_4} \, {{\cal R}}_4 \, {\cal V}_c + ....
\eea
where the dots denote the additional (flux-dependent) contributions and $g_s$ is the string coupling, while ${\cal V}_c$ denotes the compactification volume of internal Calabi Yau (CY) manifold. From now onwards, we will consider a dimensionless parameter ${\cal V}_s$ defined by: ${\cal V}_c = {\cal V}_s \, (\alpha^\prime)^3$, as the string-frame compactification volume, which in our convention is given by~\cite{McAllister:2008hb}:
\bea
\label{eq:volS}
{\cal V}_s = \frac{(2 \pi)^6}{3 !}\, \kappa_{\alpha \beta \gamma} t^\alpha\, t^\beta \, t^\gamma\,,
\eea
where $t^i$'s are dimensionless parameters for volume of the two-cycles, and $\kappa_{\alpha \beta \gamma}$ are the intersection numbers. Comparing the 4D action given in (\ref{eq:10dto4d})  
with the Einstein-Hilbert 4-D-action, yields 
\bea
& & \frac{M_p^2}{2} \equiv \frac{{\cal V}_c}{(2 \pi)^7 \, (\alpha^\prime)^4 \, g_s^2} = \frac{{\cal V}_s }{(2 \pi)^7 \, g_s^2} \, m_s^2
\eea
which subsequently gives an important relationship: 
\bea\label{msmp}
%\boxed
{m_s \equiv {l_s}^{-1}\simeq \frac{g_s \, (2 \pi)^{7/2}}{ \sqrt{2 \, {\cal V}_s}} \, M_p ~. ~}
\eea

\subsection{Kaluza-Klein $m_{KK}$, and Winding modes $m_{W}$}
Considering the toroidal orientifold compactifications, the mass scales corresponding to the KK modes and winding modes are given by, see \cite{Dienes:2002ze}:
\bea
 m_{KK} \equiv \frac{1}{{R}} = \frac{m_s}{R_0} \,,~~~
 m_W \equiv  \frac{R}{{\alpha^\prime}} =  R_0 \, m_s\,,
\eea
where $R$ and ${\alpha^\prime}/R$ are the respective radii of the KK-and their T-dual winding modes, and $R = R_0 \, l_s$ for a dimensionless parameter $R_0$. In principle, the KK-modes would depend on volumes of various internal cycles in a given CY orientifold compactification, however the LKK can be estimated by the overall compactification volume, ${\cal V}_c \equiv (2 \pi R)^6$. In terms of dimensionless parameters $R_0$ and ${\cal V}_s$ satisfying $(2 \pi R_0)^6 \equiv {\cal V}_s$, we obtain the LKK mass:
\bea
\label{eq:Mkk}
%\boxed
{m_{KK} \simeq \frac{2 \, \pi}{ {{\cal V}_s^{1/6}}} \, m_s  \simeq \frac{g_s \, (2 \pi)^{9/2}}{ \sqrt{2} \, \, \, \, {\cal V}_s^{2/3}} \, M_p ~. ~}
\eea
Other KK-modes along with the winding modes are heavier than $m_{KK}$,  and for the validity of an EFT description, we would need $R_0 > 1$.

\subsection{Scale of inflation $(V_{\rm inf})^{1/4}$ and Hubble scale ($H_{\rm inf}$)} 
The scale of inflation is determined  by the total energy density stored in the inflaton sector. For a {\it slow-roll}
inflation, the Hubble scale is solely determined by the potential energy, 
\bea
3 \, H_{\rm inf}^2 \approx \frac{V_{\rm inf}}{M_p^2}.
\eea
The observations from Planck suggest that 
the primordial perturbations are {\it adiabatic}, Gaussian, and the temperature anisotropy of CMB is given by the magnitude of the scalar power spectrum 
$P_S$~\cite{Planck:2013jfk}
\bea\label{PowerS}
& & P_S \equiv \frac{H_{\rm inf}^2}{8\, \pi^2 M_p^2\epsilon} \left(1 + ....\right)\sim 2.2 \times 10^{-9}
\eea
where $\epsilon$ is one of the two slow-roll parameters defined as
\bea
\epsilon = \frac{M_p^2}{2} \left(\frac{V_{\rm inf}'}{V_{\rm inf}}\right)^2, \quad \eta =M_p^2 \,  \left(\frac{V_{\rm inf}''}{V_{\rm inf}}\right)
\eea
where prime denotes derivative w.r.t the inflaton field, and 
dots represent slow-roll suppressed contributions. Typically, for a slow roll inflation $\epsilon,~\eta \ll 1$. The tilt in the scalar power spectrum is given by $n_s \simeq 1 + 2 \, \eta - 6 \epsilon$, and the tensor-to-scalar ratio is denoted by $r\equiv P_T/P_S$, which yields with the help of Eq.~(\ref{PowerS}),
\bea
\label{eq:HvsRatio1}
& & {H_{\rm inf} \simeq \sqrt{\frac{r}{0.1}} \times \left(3 \times 10^{-5} \right) M_p ~\,,}
\eea
and
\bea
\label{eq:HvsRatio}
& & \hskip-0.7cm (V_{\rm inf})^{1/4} \simeq \left({\frac{r}{0.1}}\right)^{1/4} \times \left(8 \times 10^{-3} \right) M_p.
\eea
As we will see later, these aforementioned relations can be considered as bridging relations for cosmological observables and stringy parameters after writing $V_{\rm inf}$ and $H_{\rm inf}$ in terms of stringy ingredients. The current data does not conclusively say whether it is a single or multi field inflation~\cite{Planck:2013jfk}, but lack of isocurvature perturbation means that whatever isocurvature fluctuations were generated during inflation must have been transferred {\it completely} into the adiabatic modes~\cite{Enqvist:2002rf}, therefore we will mainly concentrate on a single field model of inflation.
Our bounds will also be valid for those models where there exists a late time {\it dynamical attractor} for multi fields, see {\it assisted inflation}~\cite{Liddle:1998jc, Copeland:1999cs}.

\section{Reconciling  various scales of the effective four dimensional theory }
Now we utilize the CMB relations given in eq. (\ref{eq:HvsRatio1}) and eq. (\ref{eq:HvsRatio}) as a link between the stringy parameters and cosmological variables as the left-hand side of these relation can be explicitly known in a given string model of inflation. The various constraints in Eqs.~(\ref{Bound},~\ref{Bound-1}) would yield many inequalities:
\bea
m_{s} < M_p \Longrightarrow {\cal V}_s > {(2 \, \pi)^6 \, \times \pi \, g_s^2}
\eea
which is very naturally satisfied in any given setup, since from Eq.~(\ref{eq:Mkk}):
\bea
\label{eq:Mkk_Ms}
 m_{KK} < m_s \Longrightarrow {\cal V}_s > (2 \, \pi)^6\,.
\eea
In order to make all the KK-modes lighter than the stringy excitations $m_s$, not only the overall CY volume but also all the  $t^i$'s appearing in Eq.~(\ref{eq:volS}) have to be larger than unity. While performing the moduli stabilization in type IIB string compactification, sometimes it is preferred to work in the Einstein frame, where the CY volumes in two frames are related by, ${\cal V}_E$, as ${\cal V}_s \equiv g_s^{3/2} \, {\cal V}_E$. 

\subsection{Upper bound on tensor-to-scalar ratio ($r$)}
Of course, how large the compactification volume (${\cal V}_E$) should be for an EFT argument to hold good is still debatable, but the most important constraint in this regard comes from demanding that the Hubble scale is lower than KK mass scale, i.e. $H_{\rm inf} < m_{KK}$, which results in the following inequality,
\begin{equation}
\label{eq:Hinf_Mkk}
 \frac{{\cal V}_s}{(2 \, \pi)^6}<  10^6 \times \left(\frac{g_s^4}{r}\right)^{3/4} \times \left(\frac{10 {\pi}}{9}\right)^{3/4}\,,
\end{equation}
Moreover, if we demand that inflationary scale $(V_{\rm inf})^{1/4}$ is also lower than the KK mass scale, then we obtain an even stronger constraint,
\begin{equation}
\label{eq:Vinf_Mkk}
 \frac{{\cal V}_s}{(2 \, \pi)^6}<  \left(\frac{g_s^4}{r}\right)^{3/8} \times \left(1.4 \times 10^3 \right) \,,
\end{equation}
From Eqs.~(\ref{eq:Mkk_Ms},~\ref{eq:Vinf_Mkk}), we obtain
\begin{equation}
\label{eq:main1}
{1 \ll \frac{{\cal V}_s}{(2 \, \pi)^6} <  \left(\frac{g_s^4}{r}\right)^{3/8} \times \left(1.4 \times 10^3 \right).}
\end{equation}
This is an interesting inequality which relates stringy parameters $({\cal V}_s, g_s)$ with cosmological observable $r$. Considering $$v_s = \frac{{\cal V}_s}{(2 \pi)^6} \times 10^{-3},$$ we obtain a region of parameter space $(v_s,~g_s,~r)$ as shown in Fig.1 (shaded region) which is allowed by the current cosmological observational bounds arising from Eqs.~(\ref{eq:HvsRatio1},~\ref{eq:HvsRatio}). 
\begin{figure}[H]
\centering
\includegraphics[scale=0.90]{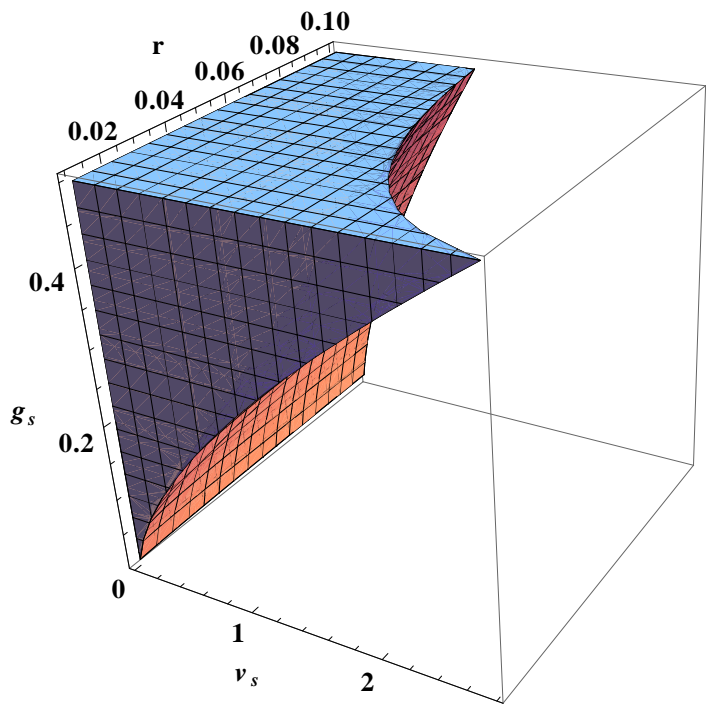}
\caption{Region plot of inequality given by Eq. (\ref{eq:main1}) showing the allowed region (shaded) in the parameter space $(v_s, g_s, r)$.}
\label{plot1}
\end{figure}
Moreover, the figure \ref{plot1} shows that a great portion of the total parameter space is already forbidden, i.e. the allowed region is given by the shaded region. Considering the latest analysis of  BICEP2/Keck Array and PLANCK results \cite{Ade:2015tva,Ade:2015lrj}, one can still hope that in future large tensor-to-scalar ratio could still be detectable, such as  $r\simeq {\cal O}(0.05)$, or so. If this were the case, we would be able to provide what are the allowed regions for $ (v_s, g_s)$, see Fig. 2. Pinning down individual values of  $g_s$ and $v_s$ would require constraints arising from  $P_S$ and $n_s$, and also the slow-roll parameters such as $\epsilon,~\eta$, therefore require some more model dependent input.
\begin{figure}[H]
\centering
\includegraphics[scale=0.95]{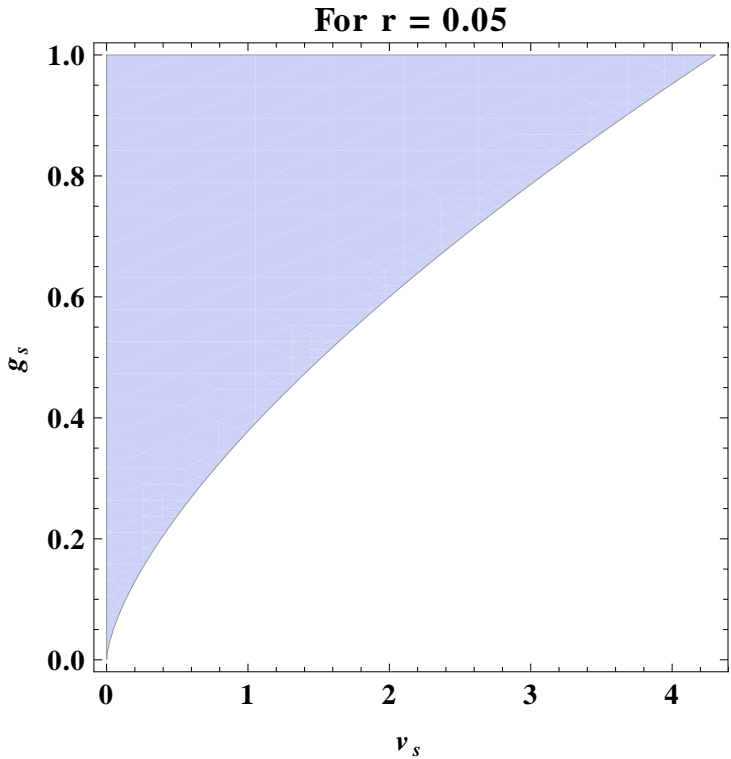}
\caption{Region of $(v_s,~g_s)$-parameter space allowed (shaded) for $r = 0.05$.}
\label{plot2}
\end{figure}
An immediate observation from Fig.\ref{plot2} is the fact that there are some particular indirect constraints on the flux landscape as well. For example, say if in future the value of $r$ is detected to be around $r \sim 0.05$, then while performing the ``two-step moduli stabilization" as in KKLT or LVS models, one has to keep flux choices such that the stabilized values of $g_s$ and ${\cal V}_s$, both remain in the allowed (shaded) region.

Let us note that by considering the relation ${\cal V}_s \equiv g_s^{3/2} \, {\cal V}_E$, one can completely eliminate the string coupling ($g_s$) dependence from the second inequality in Eq. (\ref{eq:main1}), which leads to,
\begin{equation}
\label{eq:rbound1}
g_s \gg {r}^{1/4} \times \left(0.803 \times 10^{-2} \right) \,
\end{equation}
and
\begin{equation}
\label{eq:rbound2}
 \frac{{\cal V}_E}{(2 \, \pi)^6} \ll \left(\frac{1}{r}\right)^{3/8} \times \left(1.4 \times 10^3 \right)\,, 
\end{equation}
or, equivalently
\begin{equation}\label{rmax}
 r \ll \left(2.4 \times 10^8\right) \, \, \times \left(\frac{(2\, \pi)^{16}}{{\cal V}_E^{8/3}} \right) \equiv r_{max} , 
\end{equation}
The above constraints are interesting because the two stringy parameters, ${\cal V}_E$ and $g_s$, are now constrained entirely through $r$. Considering $$v_e = \frac{{\cal V}_E}{(2 \pi)^6} \times 10^{-3}$$ the allowed shaded region for $(v_e,~r)$  are now shown in Fig.3.
\begin{figure}[H]
\centering
\includegraphics[scale=1.0]{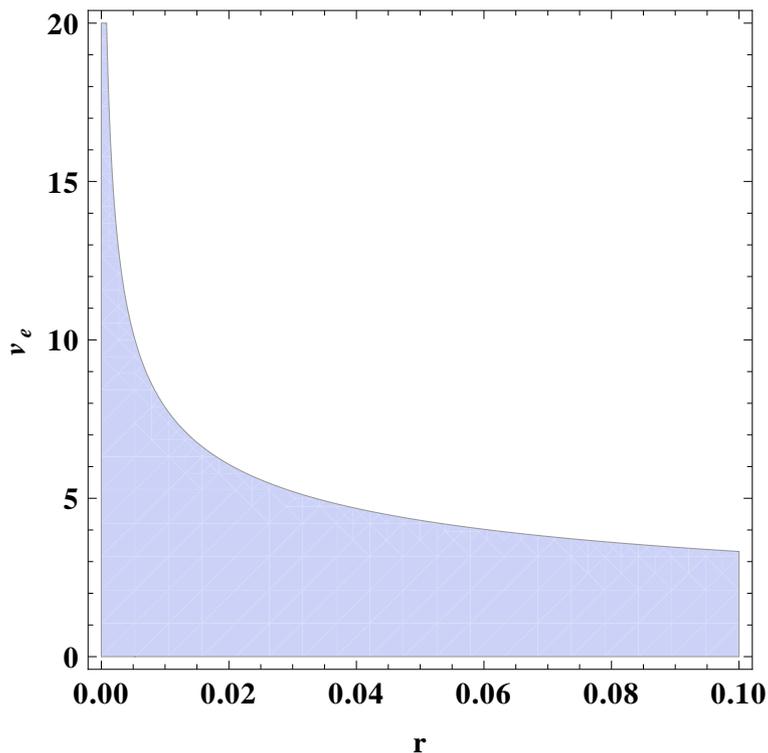}
\caption{Allowed shaded region for $(v_e,~r)$ bounded by Eq.~(\ref{rmax}).}
\label{plot3}
\end{figure}
For a numerical estimate, if $r= 0.05$, then ${\cal V}_E/(2 \pi)^6 \ll 4.3 \times 10^3$, and Eq.~(\ref{eq:rbound1}) implies that $g_s  \gg 3.8 \times 10^{-3}$. This bound on $g_s$ can be easily satisfied in any string model, which is concordant to our assumptions of validity of EFT. In a realistic scenario, the bound on $g_s$ can  be further tightened as follows. 

%\subsubsection*{Comments on control over higher order $\alpha^\prime$ and string-loop ($g_s$) corrections}
For any realistic model within string theory, one must have control over an infinite series of $\alpha^\prime$-corrections \cite{Becker:2002nn,Ciupke:2015msa} and string loop-corrections \cite{Berg:2005ja,Berg:2007wt,Cicoli:2007xp}, which are typically suppressed in powers of internal Calabi Yau volume. For the sake of illustration, if we demand ${\cal V}_s/(2 \pi)^6 > 100$ (instead of considering ${\cal V}_s/(2 \pi)^6 > 1$) in Eq.~(\ref{eq:main1}), without any additional loss of generality, we improve the bound given in Eq. (\ref{eq:rbound1}) to be
\begin{equation}
g_s > 0.17 \times r^{1/4}.
\end{equation}
This constraint implies that, for $r = 0.05$, string coupling should be fairly large, i.e. $g_s> 0.08$ as shown in the shaded region of  Fig.4.
\begin{figure}[h]
\centering
\includegraphics[scale=1.0]{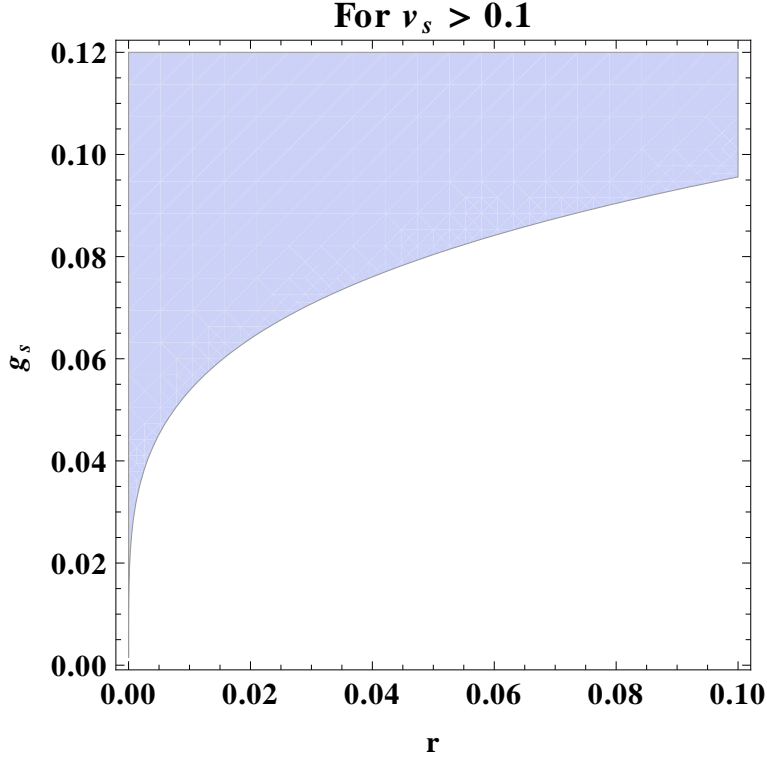}
\caption{Allowed lower bound on string coupling $g_s$ plotted for $r$ assuming that ${\cal V}_s/(2 \pi)^6 > 100$, i.e. $v_s >0.1$ in order to suppress 
$\alpha'$ and string loop corrections.}
\label{plot4}
\end{figure}

\subsection{Lower bound on tensor-to-scalar ratio ($r$)}
It is equally important to ask if we were able to place a  lower bound on $r$. In particular, the value of $r$ could be very small and still one can satisfy all other cosmological constraints, even exciting the right thermal degrees of freedom~\cite{Allahverdi:2006iq,Allahverdi:2006we}. In principle, we might be able to impose 
another simple constraint arising from the 4-D supersymmetric (SUSY) partner of graviton, i.e., gravitino, whose mass, $m_{3/2}$, must be below the LKK mass scale, 
\begin{equation}
\label{gravitino}
m_{3/2} < m_{KK} < m_s\,.
\end{equation}
In order to understand the bound arising from $m_{3/2}$, 
we would need to understand the 4-D effective potential obtained from dimensional reduction,
which has three building blocks; namely the K\"ahler potential ($K$), the superpotential ($W$), and the gauge-kinetic function (${\cal G}$). 
The F-term contribution to the scalar potential can be computed as,
\begin{equation}
V \equiv e^{K/M_p^2} \left[K^{I\ov J} \, (D_I W) ({\ov D}_{\ov J} {\ov W}) -3\, \frac{|W|^2}{M_p^2}\right]\,,
\end{equation}
where the 4-D gravitino mass is given by
\begin{equation}
 m_{3/2} \simeq \frac{g_s^2 \, e^{\frac{K_{cs}}{2}}\, (2\, \pi)^6\, |W_0|}{\sqrt{4 \pi}{\cal V}_s} \, M_p \, \simeq \frac{g_s \, e^{\frac{K_{cs}}{2}}\, (2\, \pi)^2 \, |W_0|}{\sqrt{\cal V}_s} \, m_s\,,
\end{equation}
where we have used Eq.~(\ref{msmp}), and $K_{cs}$ denotes the complex structure moduli part of the K\"ahler potential, and 
$W_0$ is the normalised tree level flux superpotential~\cite{Baumann:2014nda, Conlon:2006gv}. Now, imposing Eq.~(\ref{gravitino}), we get:
\begin{equation}
\label{eq:m32_Mkk}
 m_{3/2} <m_{KK} < m_s 
\Longrightarrow \frac{g_s}{2 \pi} \, e^{\frac{K_{cs}}{2}} |W_0| < \frac{{\cal V}_s^{1/3}}{(2 \pi)^2}\,.
\end{equation}
This condition is consistent with those in \cite{Cicoli:2013swa}. Now we may consider two viable possibilities:
\begin{itemize}
\item{$m_{3/2} \geq H_{\rm inf}$ has been considered by many, see for instance, Ref.~\cite{Kallosh:2004yh,Kallosh:2007wm,Ibanez:2014kia,Ibanez:2014zsa,Conlon:2008cj}, where we obtain:
\begin{eqnarray}
\label{eq:m32H_inf1}
&& \hskip-0.5cm r \leq  \left(\frac{(2\pi)^{11}}{18} \times 10^9\right) \,\times\left(\frac{g_s^4 \, e^{K_{cs}} \, |W_0|^2}{{\cal V}_s^2}\right) \nonumber\\
& &  \, \, \ll \frac{10\, \pi}{9} \times 10^8 \, \, \times \left(\frac{(2\, \pi)^8}{{\cal V}_E^{4/3}} \right)\,,
\end{eqnarray}
where Eq.~(\ref{eq:m32_Mkk}) and ${\cal V}_s = g_s^{3/2}\, {\cal V}_E$ have been used in the second step. Note that $H_{\rm inf} \leq m_{3/2}$ (along with $m_{3/2} < m_{KK}$) does not introduce any new constraint and falls within the bound in Eq.~(\ref{eq:rbound2}), already obtained in the limit: $(V_{\rm inf})^{1/4} < m_{KK}$.}

\item{$m_{3/2} \ll H_{\rm inf}$, example of this class has been considered earlier in Ref.~\cite{Cicoli:2012sz}. We obtain,
\begin{equation}
\label{eq:m32H_inf2}
%& & \hskip-0.75cm m_{3/2} \ll H_{\rm inf} \nonumber\\
%& & \hskip-0.75cm \Longrightarrow r \gg  
\frac{(2\pi)^{11}}{18} \times 10^9 \,\times\left(\frac{g_s \, e^{K_{cs}} \, |W_0|^2}{{\cal V}_E^2}\right) \equiv  r_{min}  \ll r\,.
\end{equation}
}
\end{itemize}
Now combining Eqs.~(\ref{eq:rbound2}) and (\ref{eq:m32H_inf2}), one obtains 
\begin{equation}
\label{eq:main2}
\frac{10^{9}}{(36\, \pi)}\cdot\left(\frac{g_s \, e^{K_{cs}}|W_0|^2}{{\cal V}_E^2/(2\pi)^{12}}\right) \ll r \ll  \, \frac{\left(2.4 \times 10^8\right)}{{\cal V}_E^{8/3}/(2\, \pi)^{16}}.
\end{equation}
The above bounds suggest that weaker the string coupling as well as larger the internal volume is, smaller is the value of $r$. Therefore, smaller value of $r$ is more natural  to realise in a setup developed in the framework of large volume scenarios (LVS)~\cite{Balasubramanian:2005zx}.  

The upper bound on $r$, see Eq.~(\ref{eq:main2}), should always be satisfied in a realistic 
model of inflation, where the potential is flat enough to give rise to $50-60$ e-foldings of inflation. The lower bound on $r$ depends on our assumption that $m_{3/2} \ll H_{\rm inf}$ holds true. This need not to be true always, in which case we will not have any strict lower bound on $r$. 

\subsection{Correlation among the three sources of upper bounds on $r$}
Here, it is important to mention that although we have derived some very interesting inequalities relating the two stringy parameters (${\cal V}_E, g_s$) and a cosmological observable ($r$) via reconciling various scales which would possibly appear in a given string-inspired model of inflation, however we did not fix the value of any of these scales, and so is the case for the parameters ${\cal V}_E, g_s$ and $r$. Now, we have three kinds of upper bounds on $r$ originated from the sources described as under,
\begin{itemize}
\item{Assuming the slow-roll condition $\epsilon \ll 1$, which is accompanied by the relation $r = 16 \, \epsilon$, shows that
\bea
& & r = 16 \, \epsilon << 16\, = r_{max}^{(1)}. \nonumber
\eea
}
\item{ The best`experimental bound' on $r$  is given by \cite{Array:2015xqh},
\bea
& & r < 0.071= r_{max}^{(2)} \nonumber
\eea}
\item{The third possibility for an upper bound on $r$ (say `theoretical bound'), is coming from demanding  $(V_{\rm inf})^{1/4}< m_{KK}$, and results in our constraint in eq. (\ref{eq:rbound2}) which reads as,
\begin{equation}
 r \ll \left(2.4 \times 10^8\right) \, \, \times \left(\frac{(2\, \pi)^{16}}{{\cal V}_E^{8/3}} \right) \equiv r_{max} , \nonumber
\end{equation}}
\end{itemize}
It would be a relevant question to ask how our upper bound on $r$ characterised via $r_{max}$ is correlated with the first two possibilities, i.e. which amongst the three $\{r_{max}^{(1)}, r_{max}^{(2)}, r_{max} \}$ is the smallest value to be relevant as an `effective' upper bound. Let us discuss the following two possibilities,
\begin{itemize}
\item{Case-I: the theoretical bound being stronger than the experimental bound\\

Although, there is no fundamental reason to argue why our theoretical bound should be stronger than the experimental bound (as well as the relatively  weaker slow-roll bound) on $r$, however let us consider a situation when $r_{max} < r_{max}^{(2)} < r_{max}^{(1)}$. A simple comparison shows that our constraint in eq. (\ref{eq:rbound2}) will be stronger only when the followings hold,
\bea
\label{eq:slowrollr}
& & r_{max} < r_{max}^{(1)} \, \, \, \Longrightarrow \frac{{\cal V}_E}{(2 \pi)^6} > 490  \, \quad \& \quad r_{max} < r_{max}^{(2)} \, \, \, \Longrightarrow \frac{{\cal V}_E}{(2 \pi)^6} > 3.2 \times 10^3 \,. \,  \nonumber
\eea}
\item{Case-II: the experimental bound being stronger than the theoretical bound.
For this case, we have the following restrictions on the value of Einstein frame Calabi Yau volume,
\bea
\label{eq:slowrollr}
& & r_{max} > r_{max}^{(1)} \, \, \, \Longrightarrow \frac{{\cal V}_E}{(2 \pi)^6} < 490  \, \quad \& \quad r_{max} > r_{max}^{(2)} \, \, \, \Longrightarrow \frac{{\cal V}_E}{(2 \pi)^6} < 3.2 \times 10^3 \,. \,  \nonumber
\eea
}
\end{itemize}
These simple estimates show that demanding our theoretical bound to be stronger than the experimental bound results in a lower bound on the internal CY volume, while when experimental bounds are stronger than the theoretical bound, we have an upper bound on the internal CY volume.  However, given that it is better to have a large volume for having the more likely consistency within the requirement of an EFT description to be valid, i.e. regarding the control over $\alpha^\prime$- and string loop-corrections, this makes the first case more pertinent. 

Further, under these circumstances, we have a relatively tight allowed window for the CY volume modulus. These are translated into having the following inequality,
\bea
& & 3.2 \times 10^3 < \frac{{\cal V}_E}{(2 \, \pi)^6} \ll \left(\frac{1}{r}\right)^{3/8} \times \left(1.4 \times 10^3 \right)
\eea
where the two bounds coincide at $r = 0.11$. For a smaller value of $r$, a wider range of allowed values of CY volume becomes available, e.g. for $r =0.01$, the allowed values are in the range: $3.2 \times 10^3 < \frac{{\cal V}_E}{(2 \, \pi)^6} \ll 7.9 \times 10^3$. Similarly, for Starobinsky-type inflationary realization (e.g. Fibre inflation) within type IIB orietifolds which realizes $r \sim 0.005$, one needs to ensure $3.2 \times 10^3 < \frac{{\cal V}_E}{(2 \, \pi)^6} \ll 1.0 \times 10^4$, where the upper bound follows straight from demanding  $(V_{\rm inf})^{1/4}< m_{KK}$, and it is quite generic, while the lower bound comes from assuming that our upper bound is stronger than the current experimental upper bound $r < 0.11$. 

\subsection{Testing the bounds for an explicit class of model}

Let us  now consider a simple toy model example of inflation which can in principle produce large value of $r$, 
and which satisfies all the observed CMB data. It is based on the Kim-Nilles-Peloso (KNP)-mechanism \cite{Kim:2004rp} of aligned natural inflation, see also \cite{Choi:2014rja, Ben-Dayan:2014zsa, Long:2014dta, Gao:2014uha, Ben-Dayan:2014lca, Tye:2014tja}. One can embed KNP-type aligned natural inflation with various RR axions or its combinations, with a multi-racetrack superpotential~\cite{Long:2014dta,Gao:2014uha,Ben-Dayan:2014lca}, for which the potential is given by a single field inflation with the help of two sub-Planckian axionic VEVs, 
\begin{equation}
\label{eq:Vaxion4}
V(\psi) = \Lambda_0 \left(1 - \cos\left[\frac{\psi}{2 \pi \, f_{\rm eff}}\right]  \right)~,~\,
\end{equation}
where
\begin{equation}
 \psi = \frac{n_2 \, f_1 \, \phi_1 - n_1 \, f_2 \, \phi_2}{\sqrt{n_1^2 \, f_2^2 +\, n_2^2 \, f_1^2}}~, ~ f_{\rm eff} = \frac{\sqrt{n_1^2 \, f_2^2 + n_2^2 \, f_1^2}}{|n_1 \, m_2 - n_2 \, m_1|}. 
\end{equation}
Here, the two fields $\phi_a$'s  are canonically normalised stringy RR axions (say $c_a$) with $\phi_a \equiv  c_a \, f_{a}$ and their decay constant $f_a$'s depend on the model dependent parameters, such as volume of the internal manifold, string coupling, see~Refs.~\cite{Baumann:2014nda, McAllister:2008hb}. Further, $n_i$'s and $m_i$'s can be written as ${2\pi h_i}/{N_i}$, where $N_i$'s are rank of the gauge groups involved via non-perturbative superpotential, while $h_i$'s can be integer quantities such as magnetic flux quanta \cite{Gao:2014uha, Ben-Dayan:2014lca}, or winding numbers \cite{Long:2014dta}. Irrespective of the details,  we can illustrate the constraints by recalling that the potential energy density of the inflaton in the KNP model is given by~\cite{Gao:2014uha, Ben-Dayan:2014lca}
\begin{equation}
\label{eq:expr-1}
 \Lambda_0 \simeq \frac{g_s}{8 \pi} \, \frac{e^{K_{cs}} \, (2\pi)^{12} \, \,  |W_0|^2}{{\cal V}_E^2} \,   {\cal F}\, \, \simeq ~~ 4.1 \times 10^{-8} \, r \, ,\,
\end{equation}
where ${\cal F}$ is a multiplicative factor appearing as a measure of the axionic-shift-symmetry breaking. We have used, Eq. (\ref{eq:HvsRatio}) along with $ \Lambda_0 \simeq \left(3 \, H_{\rm inf}^2 \, \, M_p^2 \right)$ in the second step. For large volume models such as~\cite{Gao:2014uha, Ben-Dayan:2014lca}, after taking care of normalisation factors appropriately, the multiplicative factor ${\cal F}$ is simply given by ${\cal F} \sim (2\pi)^6 \, \delta/{\cal V}_E$, where $\delta \leq 1$ is a model dependent parameter. 
%Notice that, because of different conventions via $l_s = \sqrt{\alpha^\prime}$, one has additional factor of $(2 \pi)^6$ being rooted in (\ref{eq:volS}). 
Now, further imposing our constraint Eq.~(\ref{eq:m32_Mkk}) in this class of model yields,
\begin{equation}
\label{eq:expr-2}
 r \simeq {2.4 \times 10^7} \, \Lambda_0 \ll  \left(3.8 \times 10^7\right) \frac{(2\pi)^{14}}{{\cal V}_E^{7/3}} \, \times \delta ~.
\end{equation} 
Note that Eq~(\ref{eq:expr-2}) is compatible with our model independent upper bound on $r$ given in Eq.~(\ref{eq:main2}). In fact, for $\delta < 6/{({\cal V}_E/(2\pi)^6})^{1/3}$, which could be a consistent requirement for maintaining mass-hierarchy between inflaton and the heavier moduli/axions, the bound given in Eq.~(\ref{eq:expr-2}) is even stronger than the model independent bound of Eq.~(\ref{eq:rbound2}). For numerical estimates, if we take $\delta \simeq 0.1$, then for $r \simeq \{0.1,~ 0.05\}$, one gets ${\cal V}_E/(2 \pi)^6 < \{1772,~2385\}$, which  satisfies 
our model independent bound in Eq.~(\ref{eq:rbound2}). 

Finally, let us present the subsequent constraints on the Einstein frame Calabi Yau volumes needed for some specific models which have been developed in the context of type IIB orientifolds constructions.  Without going through all the details, let us present a list of models which has been mostly taken from the reviews presented in \cite{Burgess:2013sla},
\begin{table}[H]
  \centering
 \begin{tabular}{|c||c|c||c|}
\hline
& & &\\
Inflatinary Models & $n_s$  & $r$  & ${\cal V}_E$ \\
& & &\\
  \hline
& & &\\
Higgs-otic inflation \cite{Ibanez:2014swa} & $0.966 \leq n_s \leq 0.972$ & $0.080\leq r \leq 0.098$ & ${\cal V}_E \ll 3.35\times 10^3$ \\
Axion monodromy \cite{McAllister:2008hb,Flauger:2009ab} & $0.97 \leq n_s \leq 0.98$ & $0.04\leq r \leq 0.07$ & ${\cal V}_E \ll 3.80\times 10^3$ \\
Fibre inflation \cite{Cicoli:2008gp,Burgess:2016owb} & $0.965 \leq n_s \leq 0.97$ & $0.005\leq r \leq 0.007$ & ${\cal V}_E \ll 8.90 \times 10^4$ \\
N-flation \cite{Dimopoulos:2005ac,Grimm:2007hs} & $0.93 \leq n_s \leq 0.95$ & $r \leq 10^{-3}$ & ${\cal V}_E \ll 1.87 \times 10^4$ \\
Poly-inst. Inflation \cite{Cicoli:2011ct,Blumenhagen:2012ue,Lust:2013kt}& $0.95 \leq n_s \leq 0.97$ & $r \leq 10^{-5}$ & ${\cal V}_E \ll 1.05 \times 10^5$ \\
$D3/\ov{D3}$-Brane inflation \cite{Kachru:2003sx,Dvali:2001fw} & $0.966 \leq n_s \leq 0.972$ & $r \leq 10^{-5}$ & ${\cal V}_E \ll 1.05 \times 10^5$ \\
$D3/D7$-Brane inflation \cite{Dasgupta:2004dw,Haack:2008yb} & $0.95 \leq n_s \leq 0.97$ & $10^{-12}\leq r \leq 10^{-5}$ & ${\cal V}_E \ll 1.05 \times 10^5$ \\
Inflection point inflation \cite{Linde:2007jn} & $0.92 \leq n_s \leq 0.93$ & $r \leq 10^{-6}$ & ${\cal V}_E \ll 2.49 \times 10^5$ \\
DBI inflation \cite{Lidsey:2007gq} & $0.93 \leq n_s \leq 0.93$ & $r \leq 10^{-7}$ & ${\cal V}_E \ll 5.90 \times 10^5$ \\
Racetrack inflation \cite{BlancoPillado:2004ns,BlancoPillado:2006he} & $0.95 \leq n_s \leq 0.96$ & $r \leq 10^{-8}$ & ${\cal V}_E \ll 1.40 \times 10^6$ \\
Blow-up inflation \cite{Conlon:2005jm} & $0.96 \leq n_s \leq 0.967$ & $r \leq 10^{-10}$ & ${\cal V}_E \ll 7.87 \times 10^6$ \\
Wilson line inflation \cite{Avgoustidis:2006zp} & $0.96 \leq n_s \leq 0.97$ & $r \leq 10^{-10}$ & ${\cal V}_E \ll 7.87 \times 10^6$ \\
& & &\\
  \hline
  \end{tabular}
  \caption{Some typical values for the upper bound on the Einstein frame CY volume (${\cal V}_E$) corresponding to the maximum vales of the tensor-to-scalar ratio $r$ realized via demanding $50 < N_e < 60$ e-foldings in a set of inflationary models realized within string framework.}
 \end{table}

\section{Conclusions}
As we conclude, let us point out that a {\it sufficient} large volume of the internal CY is must in a given inflationary setup in order to have protection against various (un-)known $\alpha^\prime$ and $g_s$ corrections, as  the EFT description in a given background geometry can be trusted as long as $(V_{\rm inf})^{1/4}< m_{KK}$, and/or $H_{\rm inf} < m_{KK}$. These inequalities along with our bound Eqs.~(\ref{eq:rbound1})-(\ref{eq:rbound2}) should always be satisfied (within all the models where warping effects are negligible), and our inequalities should serves as a guiding principle for building inflationary models in (type IIB) superstring theory framework, which can explain the CMB data. In future, if we can can ascertain the value of $r$ to high accuracy, we should be indeed able to pin down some of the key stringy parameters.

\section*{Acknowledgments}
AM is supported by STFC grant ST/J000418/1. PS was supported by the Compagnia di San Paolo contract ÒModern Application of String TheoryÓ (MAST) TO-Call3-2012-0088.

%\newpage
%\bibliography{references}
\bibliographystyle{utphys}
\bibliography{reference}

\providecommand{\href}[2]{#2}\begingroup\raggedright\begin{thebibliography}{10}

\bibitem{Planck:2013jfk}
{\bf Planck} Collaboration, P.~A.~R. Ade {\em et al.}, ``{Planck 2013 results.
  XXII. Constraints on inflation},'' {\em Astron. Astrophys.} {\bf 571} (2014)
  A22,
\href{http://www.arXiv.org/abs/1303.5082}{{\tt 1303.5082}}.
%%CITATION = ARXIV:1303.5082;%%.

\bibitem{Mazumdar:2010sa}
A.~Mazumdar and J.~Rocher, ``{Particle physics models of inflation and curvaton
  scenarios},'' {\em Phys. Rept.} {\bf 497} (2011) 85--215,
\href{http://www.arXiv.org/abs/1001.0993}{{\tt 1001.0993}}.
%%CITATION = ARXIV:1001.0993;%%.

\bibitem{Chialva:2014rla}
D.~Chialva and A.~Mazumdar, ``{Cosmological implications of quantum corrections
  and higher-derivative extension},'' {\em Mod. Phys. Lett.} {\bf A30} (2015),
  no.~03n04, 1540008,
\href{http://www.arXiv.org/abs/1405.0513}{{\tt 1405.0513}}.
%%CITATION = ARXIV:1405.0513;%%.

\bibitem{Ade:2014xna}
{\bf BICEP2} Collaboration, P.~A.~R. Ade {\em et al.}, ``{Detection of $B$-Mode
  Polarization at Degree Angular Scales by BICEP2},'' {\em Phys. Rev. Lett.}
  {\bf 112} (2014), no.~24, 241101,
\href{http://www.arXiv.org/abs/1403.3985}{{\tt 1403.3985}}.
%%CITATION = ARXIV:1403.3985;%%.

\bibitem{Ade:2015tva}
{\bf BICEP2, Planck} Collaboration, P.~Ade {\em et al.}, ``{Joint Analysis of
  BICEP2/$Keck  Array$ and $Planck$ Data},'' {\em Phys. Rev. Lett.} {\bf 114}
  (2015) 101301,
\href{http://www.arXiv.org/abs/1502.00612}{{\tt 1502.00612}}.
%%CITATION = ARXIV:1502.00612;%%.

\bibitem{Ade:2015lrj}
{\bf Planck} Collaboration, P.~A.~R. Ade {\em et al.}, ``{Planck 2015 results.
  XX. Constraints on inflation},''
\href{http://www.arXiv.org/abs/1502.02114}{{\tt 1502.02114}}.
%%CITATION = ARXIV:1502.02114;%%.

\bibitem{Array:2015xqh}
{\bf BICEP2, Keck Array} Collaboration, P.~A.~R. Ade {\em et al.}, ``{Improved
  Constraints on Cosmology and Foregrounds from BICEP2 and Keck Array Cosmic
  Microwave Background Data with Inclusion of 95 GHz Band},'' {\em Phys. Rev.
  Lett.} {\bf 116} (2016) 031302,
\href{http://www.arXiv.org/abs/1510.09217}{{\tt 1510.09217}}.
%%CITATION = ARXIV:1510.09217;%%.

\bibitem{Hotchkiss:2011gz}
S.~Hotchkiss, A.~Mazumdar, and S.~Nadathur, ``{Observable gravitational waves
  from inflation with small field excursions},'' {\em JCAP} {\bf 1202} (2012)
  008,
\href{http://www.arXiv.org/abs/1110.5389}{{\tt 1110.5389}}.
%%CITATION = ARXIV:1110.5389;%%.

\bibitem{Chatterjee:2014hna}
A.~Chatterjee and A.~Mazumdar, ``{Bound on largest $r\lesssim 0.1$ from
  sub-Planckian excursions of inflaton},'' {\em JCAP} {\bf 1501} (2015),
  no.~01, 031, \href{http://www.arXiv.org/abs/1409.4442}{{\tt 1409.4442}}.

\bibitem{Choudhury:2014sxa}
S.~Choudhury, A.~Mazumdar, and E.~Pukartas, ``{Constraining ${\cal N}=1$
  supergravity inflationary framework with non-minimal KŠhler operators},''
  {\em JHEP} {\bf 04} (2014) 077,
\href{http://www.arXiv.org/abs/1402.1227}{{\tt 1402.1227}}.
%%CITATION = ARXIV:1402.1227;%%.

\bibitem{Burgess:2013sla}
C.~P. Burgess, M.~Cicoli, and F.~Quevedo, ``{String Inflation After Planck
  2013},'' {\em JCAP} {\bf 1311} (2013) 003,
\href{http://www.arXiv.org/abs/1306.3512}{{\tt 1306.3512}}.
%%CITATION = ARXIV:1306.3512;%%.

\bibitem{Westphal:2014ana}
A.~Westphal, ``{String cosmology — Large-field inflation in string theory},''
  {\em Int. J. Mod. Phys.} {\bf A30} (2015), no.~09, 1530024,
\href{http://www.arXiv.org/abs/1409.5350}{{\tt 1409.5350}}.
%%CITATION = ARXIV:1409.5350;%%.

\bibitem{Baumann:2014nda}
D.~Baumann and L.~McAllister, {\em {Inflation and String Theory}}.
\newblock Cambridge University Press,
2015.
\newblock
%%CITATION = ARXIV:1404.2601;%%.

\bibitem{Kofman:2005yz}
L.~Kofman and P.~Yi, ``{Reheating the universe after string theory
  inflation},'' {\em Phys. Rev.} {\bf D72} (2005) 106001,
\href{http://www.arXiv.org/abs/hep-th/0507257}{{\tt hep-th/0507257}}.
%%CITATION = HEP-TH/0507257;%%.

\bibitem{Frey:2005jk}
A.~R. Frey, A.~Mazumdar, and R.~C. Myers, ``{Stringy effects during inflation
  and reheating},'' {\em Phys. Rev.} {\bf D73} (2006) 026003,
\href{http://www.arXiv.org/abs/hep-th/0508139}{{\tt hep-th/0508139}}.
%%CITATION = HEP-TH/0508139;%%.

\bibitem{Cicoli:2010ha}
M.~Cicoli and A.~Mazumdar, ``{Reheating for Closed String Inflation},'' {\em
  JCAP} {\bf 1009} (2010), no.~09, 025,
\href{http://www.arXiv.org/abs/1005.5076}{{\tt 1005.5076}}.
%%CITATION = ARXIV:1005.5076;%%.

\bibitem{Cicoli:2010yj}
M.~Cicoli and A.~Mazumdar, ``{Inflation in string theory: A Graceful exit to
  the real world},'' {\em Phys. Rev.} {\bf D83} (2011) 063527,
\href{http://www.arXiv.org/abs/1010.0941}{{\tt 1010.0941}}.
%%CITATION = ARXIV:1010.0941;%%.

\bibitem{Chialva:2012rq}
D.~Chialva, P.~S.~B. Dev, and A.~Mazumdar, ``{Multiple dark matter scenarios
  from ubiquitous stringy throats},'' {\em Phys. Rev.} {\bf D87} (2013), no.~6,
  063522,
\href{http://www.arXiv.org/abs/1211.0250}{{\tt 1211.0250}}.
%%CITATION = ARXIV:1211.0250;%%.

\bibitem{Berg:2005ja}
M.~Berg, M.~Haack, and B.~Kors, ``{String loop corrections to Kahler potentials
  in orientifolds},'' {\em JHEP} {\bf 11} (2005) 030,
\href{http://www.arXiv.org/abs/hep-th/0508043}{{\tt hep-th/0508043}}.
%%CITATION = HEP-TH/0508043;%%.

\bibitem{Berg:2007wt}
M.~Berg, M.~Haack, and E.~Pajer, ``{Jumping Through Loops: On Soft Terms from
  Large Volume Compactifications},'' {\em JHEP} {\bf 09} (2007) 031,
\href{http://www.arXiv.org/abs/0704.0737}{{\tt 0704.0737}}.
%%CITATION = ARXIV:0704.0737;%%.

\bibitem{Cicoli:2007xp}
M.~Cicoli, J.~P. Conlon, and F.~Quevedo, ``{Systematics of String Loop
  Corrections in Type IIB Calabi-Yau Flux Compactifications},'' {\em JHEP} {\bf
  01} (2008) 052,
\href{http://www.arXiv.org/abs/0708.1873}{{\tt 0708.1873}}.
%%CITATION = ARXIV:0708.1873;%%.

\bibitem{Becker:2002nn}
K.~Becker, M.~Becker, M.~Haack, and J.~Louis, ``{Supersymmetry breaking and
  alpha-prime corrections to flux induced potentials},'' {\em JHEP} {\bf 06}
  (2002) 060,
\href{http://www.arXiv.org/abs/hep-th/0204254}{{\tt hep-th/0204254}}.
%%CITATION = HEP-TH/0204254;%%.

\bibitem{Ciupke:2015msa}
D.~Ciupke, J.~Louis, and A.~Westphal, ``{Higher-Derivative Supergravity and
  Moduli Stabilization},'' {\em JHEP} {\bf 10} (2015) 094,
\href{http://www.arXiv.org/abs/1505.03092}{{\tt 1505.03092}}.
%%CITATION = ARXIV:1505.03092;%%.

\bibitem{Ciupke:2016agp}
D.~Ciupke, ``{Scalar Potential from Higher Derivative $\mathcal{N} = 1$
  Superspace},''
\href{http://www.arXiv.org/abs/1605.00651}{{\tt 1605.00651}}.
%%CITATION = ARXIV:1605.00651;%%.

\bibitem{Brandenberger:1988aj}
R.~H. Brandenberger and C.~Vafa, ``{Superstrings in the Early Universe},'' {\em
  Nucl. Phys.} {\bf B316} (1989)
391--410.
%%CITATION = NUPHA,B316,391;%%.

\bibitem{Danos:2004jz}
R.~Danos, A.~R. Frey, and A.~Mazumdar, ``{Interaction rates in string gas
  cosmology},'' {\em Phys. Rev.} {\bf D70} (2004) 106010,
\href{http://www.arXiv.org/abs/hep-th/0409162}{{\tt hep-th/0409162}}.
%%CITATION = HEP-TH/0409162;%%.

\bibitem{Kachru:2003aw}
S.~Kachru, R.~Kallosh, A.~D. Linde, and S.~P. Trivedi, ``{De Sitter vacua in
  string theory},'' {\em Phys. Rev.} {\bf D68} (2003) 046005,
\href{http://www.arXiv.org/abs/hep-th/0301240}{{\tt hep-th/0301240}}.
%%CITATION = HEP-TH/0301240;%%.

\bibitem{Balasubramanian:2005zx}
V.~Balasubramanian, P.~Berglund, J.~P. Conlon, and F.~Quevedo, ``{Systematics
  of moduli stabilisation in Calabi-Yau flux compactifications},'' {\em JHEP}
  {\bf 03} (2005) 007,
\href{http://www.arXiv.org/abs/hep-th/0502058}{{\tt hep-th/0502058}}.
%%CITATION = HEP-TH/0502058;%%.

\bibitem{McAllister:2008hb}
L.~McAllister, E.~Silverstein, and A.~Westphal, ``{Gravity Waves and Linear
  Inflation from Axion Monodromy},'' {\em Phys. Rev.} {\bf D82} (2010) 046003,
\href{http://www.arXiv.org/abs/0808.0706}{{\tt 0808.0706}}.
%%CITATION = ARXIV:0808.0706;%%.

\bibitem{Dienes:2002ze}
K.~R. Dienes and A.~Mafi, ``{Kaluza-Klein states versus winding states: Can
  both be above the string scale?},'' {\em Phys. Rev. Lett.} {\bf 89} (2002)
  171602,
\href{http://www.arXiv.org/abs/hep-ph/0207009}{{\tt hep-ph/0207009}}.
%%CITATION = HEP-PH/0207009;%%.

\bibitem{Enqvist:2002rf}
K.~Enqvist, S.~Kasuya, and A.~Mazumdar, ``{Adiabatic density perturbations and
  matter generation from the MSSM},'' {\em Phys. Rev. Lett.} {\bf 90} (2003)
  091302,
\href{http://www.arXiv.org/abs/hep-ph/0211147}{{\tt hep-ph/0211147}}.
%%CITATION = HEP-PH/0211147;%%.

\bibitem{Liddle:1998jc}
A.~R. Liddle, A.~Mazumdar, and F.~E. Schunck, ``{Assisted inflation},'' {\em
  Phys. Rev.} {\bf D58} (1998) 061301,
\href{http://www.arXiv.org/abs/astro-ph/9804177}{{\tt astro-ph/9804177}}.
%%CITATION = ASTRO-PH/9804177;%%.

\bibitem{Copeland:1999cs}
E.~J. Copeland, A.~Mazumdar, and N.~J. Nunes, ``{Generalized assisted
  inflation},'' {\em Phys. Rev.} {\bf D60} (1999) 083506,
\href{http://www.arXiv.org/abs/astro-ph/9904309}{{\tt astro-ph/9904309}}.
%%CITATION = ASTRO-PH/9904309;%%.

\bibitem{Allahverdi:2006iq}
R.~Allahverdi, K.~Enqvist, J.~Garcia-Bellido, and A.~Mazumdar, ``{Gauge
  invariant MSSM inflaton},'' {\em Phys. Rev. Lett.} {\bf 97} (2006) 191304,
\href{http://www.arXiv.org/abs/hep-ph/0605035}{{\tt hep-ph/0605035}}.
%%CITATION = HEP-PH/0605035;%%.

\bibitem{Allahverdi:2006we}
R.~Allahverdi, K.~Enqvist, J.~Garcia-Bellido, A.~Jokinen, and A.~Mazumdar,
  ``{MSSM flat direction inflation: Slow roll, stability, fine tunning and
  reheating},'' {\em JCAP} {\bf 0706} (2007) 019,
\href{http://www.arXiv.org/abs/hep-ph/0610134}{{\tt hep-ph/0610134}}.
%%CITATION = HEP-PH/0610134;%%.

\bibitem{Conlon:2006gv}
J.~P. Conlon, ``{Moduli Stabilisation and Applications in IIB String Theory},''
  {\em Fortsch. Phys.} {\bf 55} (2007) 287--422,
\href{http://www.arXiv.org/abs/hep-th/0611039}{{\tt hep-th/0611039}}.
%%CITATION = HEP-TH/0611039;%%.

\bibitem{Cicoli:2013swa}
M.~Cicoli, J.~P. Conlon, A.~Maharana, and F.~Quevedo, ``{A Note on the
  Magnitude of the Flux Superpotential},'' {\em JHEP} {\bf 01} (2014) 027,
\href{http://www.arXiv.org/abs/1310.6694}{{\tt 1310.6694}}.
%%CITATION = ARXIV:1310.6694;%%.

\bibitem{Kallosh:2004yh}
R.~Kallosh and A.~D. Linde, ``{Landscape, the scale of SUSY breaking, and
  inflation},'' {\em JHEP} {\bf 12} (2004) 004,
\href{http://www.arXiv.org/abs/hep-th/0411011}{{\tt hep-th/0411011}}.
%%CITATION = HEP-TH/0411011;%%.

\bibitem{Kallosh:2007wm}
R.~Kallosh and A.~D. Linde, ``{Testing String Theory with CMB},'' {\em JCAP}
  {\bf 0704} (2007) 017,
\href{http://www.arXiv.org/abs/0704.0647}{{\tt 0704.0647}}.
%%CITATION = ARXIV:0704.0647;%%.

\bibitem{Ibanez:2014kia}
L.~E. Ibanez and I.~Valenzuela, ``{The inflaton as an MSSM Higgs and open
  string modulus monodromy inflation},'' {\em Phys. Lett.} {\bf B736} (2014)
  226--230,
\href{http://www.arXiv.org/abs/1404.5235}{{\tt 1404.5235}}.
%%CITATION = ARXIV:1404.5235;%%.

\bibitem{Ibanez:2014zsa}
L.~E. Ibanez and I.~Valenzuela, ``{BICEP2, the Higgs Mass and the SUSY-breaking
  Scale},'' {\em Phys. Lett.} {\bf B734} (2014) 354--357,
\href{http://www.arXiv.org/abs/1403.6081}{{\tt 1403.6081}}.
%%CITATION = ARXIV:1403.6081;%%.

\bibitem{Conlon:2008cj}
J.~P. Conlon, R.~Kallosh, A.~D. Linde, and F.~Quevedo, ``{Volume Modulus
  Inflation and the Gravitino Mass Problem},'' {\em JCAP} {\bf 0809} (2008)
  011,
\href{http://www.arXiv.org/abs/0806.0809}{{\tt 0806.0809}}.
%%CITATION = ARXIV:0806.0809;%%.

\bibitem{Cicoli:2012sz}
M.~Cicoli, M.~Goodsell, and A.~Ringwald, ``{The type IIB string axiverse and
  its low-energy phenomenology},'' {\em JHEP} {\bf 10} (2012) 146,
\href{http://www.arXiv.org/abs/1206.0819}{{\tt 1206.0819}}.
%%CITATION = ARXIV:1206.0819;%%.

\bibitem{Kim:2004rp}
J.~E. Kim, H.~P. Nilles, and M.~Peloso, ``{Completing natural inflation},''
  {\em JCAP} {\bf 0501} (2005) 005,
\href{http://www.arXiv.org/abs/hep-ph/0409138}{{\tt hep-ph/0409138}}.
%%CITATION = HEP-PH/0409138;%%.

\bibitem{Choi:2014rja}
K.~Choi, H.~Kim, and S.~Yun, ``{Natural inflation with multiple sub-Planckian
  axions},'' {\em Phys. Rev.} {\bf D90} (2014) 023545,
\href{http://www.arXiv.org/abs/1404.6209}{{\tt 1404.6209}}.
%%CITATION = ARXIV:1404.6209;%%.

\bibitem{Ben-Dayan:2014zsa}
I.~Ben-Dayan, F.~G. Pedro, and A.~Westphal, ``{Hierarchical Axion Inflation},''
  {\em Phys. Rev. Lett.} {\bf 113} (2014) 261301,
\href{http://www.arXiv.org/abs/1404.7773}{{\tt 1404.7773}}.
%%CITATION = ARXIV:1404.7773;%%.

\bibitem{Long:2014dta}
C.~Long, L.~McAllister, and P.~McGuirk, ``{Aligned Natural Inflation in String
  Theory},'' {\em Phys. Rev.} {\bf D90} (2014) 023501,
\href{http://www.arXiv.org/abs/1404.7852}{{\tt 1404.7852}}.
%%CITATION = ARXIV:1404.7852;%%.

\bibitem{Gao:2014uha}
X.~Gao, T.~Li, and P.~Shukla, ``{Combining Universal and Odd RR Axions for
  Aligned Natural Inflation},'' {\em JCAP} {\bf 1410} (2014) 048,
\href{http://www.arXiv.org/abs/1406.0341}{{\tt 1406.0341}}.
%%CITATION = ARXIV:1406.0341;%%.

\bibitem{Ben-Dayan:2014lca}
I.~Ben-Dayan, F.~G. Pedro, and A.~Westphal, ``{Towards Natural Inflation in
  String Theory},'' {\em Phys. Rev.} {\bf D92} (2015), no.~2, 023515,
\href{http://www.arXiv.org/abs/1407.2562}{{\tt 1407.2562}}.
%%CITATION = ARXIV:1407.2562;%%.

\bibitem{Tye:2014tja}
S.~H.~H. Tye and S.~S.~C. Wong, ``{Helical Inflation and Cosmic Strings},''
\href{http://www.arXiv.org/abs/1404.6988}{{\tt 1404.6988}}.
%%CITATION = ARXIV:1404.6988;%%.

\bibitem{Ibanez:2014swa}
L.~E. Ibanez, F.~Marchesano, and I.~Valenzuela, ``{Higgs-otic Inflation and
  String Theory},'' {\em JHEP} {\bf 01} (2015) 128,
\href{http://www.arXiv.org/abs/1411.5380}{{\tt 1411.5380}}.
%%CITATION = ARXIV:1411.5380;%%.

\bibitem{Flauger:2009ab}
R.~Flauger, L.~McAllister, E.~Pajer, A.~Westphal, and G.~Xu, ``{Oscillations in
  the CMB from Axion Monodromy Inflation},'' {\em JCAP} {\bf 1006} (2010) 009,
\href{http://www.arXiv.org/abs/0907.2916}{{\tt 0907.2916}}.
%%CITATION = ARXIV:0907.2916;%%.

\bibitem{Cicoli:2008gp}
M.~Cicoli, C.~P. Burgess, and F.~Quevedo, ``{Fibre Inflation: Observable
  Gravity Waves from IIB String Compactifications},'' {\em JCAP} {\bf 0903}
  (2009) 013,
\href{http://www.arXiv.org/abs/0808.0691}{{\tt 0808.0691}}.
%%CITATION = ARXIV:0808.0691;%%.

\bibitem{Burgess:2016owb}
C.~P. Burgess, M.~Cicoli, S.~de~Alwis, and F.~Quevedo, ``{Robust Inflation from
  Fibrous Strings},'' {\em JCAP} {\bf 1605} (2016), no.~05, 032,
\href{http://www.arXiv.org/abs/1603.06789}{{\tt 1603.06789}}.
%%CITATION = ARXIV:1603.06789;%%.

\bibitem{Dimopoulos:2005ac}
S.~Dimopoulos, S.~Kachru, J.~McGreevy, and J.~G. Wacker, ``{N-flation},'' {\em
  JCAP} {\bf 0808} (2008) 003,
\href{http://www.arXiv.org/abs/hep-th/0507205}{{\tt hep-th/0507205}}.
%%CITATION = HEP-TH/0507205;%%.

\bibitem{Grimm:2007hs}
T.~W. Grimm, ``{Axion inflation in type II string theory},'' {\em Phys. Rev.}
  {\bf D77} (2008) 126007,
\href{http://www.arXiv.org/abs/0710.3883}{{\tt 0710.3883}}.
%%CITATION = ARXIV:0710.3883;%%.

\bibitem{Cicoli:2011ct}
M.~Cicoli, F.~G. Pedro, and G.~Tasinato, ``{Poly-instanton Inflation},'' {\em
  JCAP} {\bf 1112} (2011) 022,
\href{http://www.arXiv.org/abs/1110.6182}{{\tt 1110.6182}}.
%%CITATION = ARXIV:1110.6182;%%.

\bibitem{Blumenhagen:2012ue}
R.~Blumenhagen, X.~Gao, T.~Rahn, and P.~Shukla, ``{Moduli Stabilization and
  Inflationary Cosmology with Poly-Instantons in Type IIB Orientifolds},'' {\em
  JHEP} {\bf 11} (2012) 101,
\href{http://www.arXiv.org/abs/1208.1160}{{\tt 1208.1160}}.
%%CITATION = ARXIV:1208.1160;%%.

\bibitem{Lust:2013kt}
D.~Lüst and X.~Zhang, ``{Four Kahler Moduli Stabilisation in type IIB
  Orientifolds with K3-fibred Calabi-Yau threefold compactification},'' {\em
  JHEP} {\bf 05} (2013) 051,
\href{http://www.arXiv.org/abs/1301.7280}{{\tt 1301.7280}}.
%%CITATION = ARXIV:1301.7280;%%.

\bibitem{Kachru:2003sx}
S.~Kachru, R.~Kallosh, A.~D. Linde, J.~M. Maldacena, L.~P. McAllister, and
  S.~P. Trivedi, ``{Towards inflation in string theory},'' {\em JCAP} {\bf
  0310} (2003) 013,
\href{http://www.arXiv.org/abs/hep-th/0308055}{{\tt hep-th/0308055}}.
%%CITATION = HEP-TH/0308055;%%.

\bibitem{Dvali:2001fw}
G.~R. Dvali, Q.~Shafi, and S.~Solganik, ``{D-brane inflation},'' in {\em {4th
  European Meeting From the Planck Scale to the Electroweak Scale (Planck 2001)
  La Londe les Maures, Toulon, France, May 11-16, 2001}}.
\newblock 2001.
\newblock
\href{http://www.arXiv.org/abs/hep-th/0105203}{{\tt hep-th/0105203}}.
\newblock
%%CITATION = HEP-TH/0105203;%%.

\bibitem{Dasgupta:2004dw}
K.~Dasgupta, J.~P. Hsu, R.~Kallosh, A.~D. Linde, and M.~Zagermann, ``{D3/D7
  brane inflation and semilocal strings},'' {\em JHEP} {\bf 08} (2004) 030,
\href{http://www.arXiv.org/abs/hep-th/0405247}{{\tt hep-th/0405247}}.
%%CITATION = HEP-TH/0405247;%%.

\bibitem{Haack:2008yb}
M.~Haack, R.~Kallosh, A.~Krause, A.~D. Linde, D.~Lust, and M.~Zagermann,
  ``{Update of D3/D7-Brane Inflation on K3 x T**2/Z(2)},'' {\em Nucl. Phys.}
  {\bf B806} (2009) 103--177,
\href{http://www.arXiv.org/abs/0804.3961}{{\tt 0804.3961}}.
%%CITATION = ARXIV:0804.3961;%%.

\bibitem{Linde:2007jn}
A.~D. Linde and A.~Westphal, ``{Accidental Inflation in String Theory},'' {\em
  JCAP} {\bf 0803} (2008) 005,
\href{http://www.arXiv.org/abs/0712.1610}{{\tt 0712.1610}}.
%%CITATION = ARXIV:0712.1610;%%.

\bibitem{Lidsey:2007gq}
J.~E. Lidsey and I.~Huston, ``{Gravitational wave constraints on
  Dirac-Born-Infeld inflation},'' {\em JCAP} {\bf 0707} (2007) 002,
\href{http://www.arXiv.org/abs/0705.0240}{{\tt 0705.0240}}.
%%CITATION = ARXIV:0705.0240;%%.

\bibitem{BlancoPillado:2004ns}
J.~J. Blanco-Pillado, C.~P. Burgess, J.~M. Cline, C.~Escoda, M.~Gomez-Reino,
  R.~Kallosh, A.~D. Linde, and F.~Quevedo, ``{Racetrack inflation},'' {\em
  JHEP} {\bf 11} (2004) 063,
\href{http://www.arXiv.org/abs/hep-th/0406230}{{\tt hep-th/0406230}}.
%%CITATION = HEP-TH/0406230;%%.

\bibitem{BlancoPillado:2006he}
J.~J. Blanco-Pillado, C.~P. Burgess, J.~M. Cline, C.~Escoda, M.~Gomez-Reino,
  R.~Kallosh, A.~D. Linde, and F.~Quevedo, ``{Inflating in a better
  racetrack},'' {\em JHEP} {\bf 09} (2006) 002,
\href{http://www.arXiv.org/abs/hep-th/0603129}{{\tt hep-th/0603129}}.
%%CITATION = HEP-TH/0603129;%%.

\bibitem{Conlon:2005jm}
J.~P. Conlon and F.~Quevedo, ``{Kahler moduli inflation},'' {\em JHEP} {\bf 01}
  (2006) 146,
\href{http://www.arXiv.org/abs/hep-th/0509012}{{\tt hep-th/0509012}}.
%%CITATION = HEP-TH/0509012;%%.

\bibitem{Avgoustidis:2006zp}
A.~Avgoustidis, D.~Cremades, and F.~Quevedo, ``{Wilson line inflation},'' {\em
  Gen. Rel. Grav.} {\bf 39} (2007) 1203--1234,
\href{http://www.arXiv.org/abs/hep-th/0606031}{{\tt hep-th/0606031}}.
%%CITATION = HEP-TH/0606031;%%.

\end{thebibliography}\endgroup

\end{document}